# Phase-space analysis of a two-section InP laser as an all-optical spiking neuron: dependency on control and design parameters

**Lukas Puts[1], Daan Lenstra[1], Kevin Williams[1], and Weiming Yao[1]**

[1] Photonic Integration Group, Eindhoven Hendrik Casimir Institute (EHCI), P.O. Box 513, 5600 MB Eindhoven, the Netherlands

E-mail: l.puts@tue.nl

## Abstract

Using a rate-equation model we numerically evaluate the carrier concentration and photon number in an integrated two-section semiconductor laser, and analyse its dynamics in three-dimensional phase space. The simulation comprises compact model descriptions extracted from a commercially-available generic InP technology platform, allowing us to model an applied reverse-bias voltage to the saturable absorber. We use the model to study the influence of the injected gain current, reverse-bias voltage, and cavity mirror reflectivity on the excitable operation state, which is the operation mode desired for the laser to act as an all-optical integrated neuron. We show in phase-space that our model is capable of demonstrating four different operation modes, i.e. cw, self-pulsating and an on-set and excitable mode under optical pulse injection. In addition, we show that lowering the reflectivity of one of the cavity mirrors greatly enhances the control parameter space for excitable operation, enabling more relaxed operation parameter control and lower power consumption of an integrated two-section laser neuron.

## 1. Introduction

Electronic processors in form of CPUs and GPUs are widely used today to implement powerful machine learning systems based on artificial neural networks (ANN). Since these processors are based on the Von-Neumann computer architecture, data are processed sequentially. Inspired by the human brain, an ANN in contrast consists of a large amount of parallel artificial neurons and a large number of weighted interconnections representing biological synapses. The vector-matrix multiplication performed in ANNs is a parallel computational process, which limits the efficiency of ANNs implemented on traditional electronic processors. In addition, the computational power needed for novel ANNs currently outgrows the computational power delivered by traditional hardware [1], [2]. Thus, novel hardware designed specifically to implement ANNs on a chip has been the subject of research for a while.

A specific type of ANN that mimics the dynamics observed in biological neurons is the spiking neural network (SNN). In the human brain, neurons are connected using dendrites and axons. An axon conveys the action potential generated by a neuron to consecutive neurons, which is essentially an all-or-nothing spiking signal. With the condition that a spike is sufficiently strong, consecutive neurons are triggered to generate a response spike [3]–[5], which is referred to as 'excitability'. In addition to single spike excitation, biological neurons exhibit other dynamical features, such as periodic spiking and bursting. A detailed analysis of the dynamics of biological neurons can be found in [6].

The rich dynamics observed in biological neurons has been exploited in SNNs to tackle various machine learning tasks [7], [8]. Due to the potential lower energy consumption, large bandwidth, low crosstalk and extremely fast spiking rates, integrated photonics is a promising candidate to explore the implementation of artificial neural networks [9]. The spiking mechanism can be implemented in photonics using two-section lasers. For example, by adding a Thulium-doped fibre saturable absorber in an Erbium-doped fibre laser, an excitable laser was demonstrated [10]. Similarly, integrated lasers such as vertical-cavity surface emitting lasers (VCSELs) [4], [11] and districted feedback (DFB) [12] lasers have shown great potential as optical neurons.

The dynamics of two-section lasers under optical pulse [13] or noise [14] injection as well as subject to perturbations in gain pump power [15] have been studied extensively before. Excitable lasers with delayed optical feedback have been another topic of interest [16]–[18]. These studies focus on identifying stability regions, period orbits in phase space and triggering. However, these extensive mathematical studies are based on the dimensionless Yamada model [19] and thus do not allow to directly study the influence of design and control parameters such as injection current, absorber voltage, and mirror reflectivities on the operation of a concrete laser design that can be realised in a photonic integration process. Secondly, the Yamada model is too simple with regard to certain aspects of gain and absorption. For example, device and technology specific parameters such as reverse-voltage operation and a voltage dependent carrier lifetime in the absorber are not considered in the Yamada model but are present in real laser devices.

Previously, we have studied the laser operation regimes of a two-section laser under optical pulse injection. Based on laser design parameter changes [20] and different absorber carrier lifetimes [21], the laser operation regimes have been identified. Another study discusses the excitability of two-section lasers using a lumped-cavity description [22]. Due to the rich carrier dynamics, two-section lasers exhibit different operation modes such as a self-pulsating, cw, on-set and excitable mode. So far, these studies are based on time-trace simulations without dynamical analysis in phase space, the rich dynamics exhibited by these lasers is still to be explored.

In this work, we are using the material and platform-specific rate-equation model reported in [22] and combine time-trace analysis with phase space investigation in order to study the effect of design and control parameters on laser excitability. This model has been matched to the gain building block performance on the commercial InP platform [23] using compact models for a voltage dependent absorber carrier lifetime and transparency carrier density. The model has proven to yield qualitative results that agree with experimental observations, such as a hysteresis and bistable operation [22]. This allows us to study the behaviour of gain and absorber carrier densities and photon number concentrations of realistic devices that could be taped-out to fabrication. Specifically, we will show how laser control parameters, i.e.. the forward gain current and a reverse-bias absorber voltage change the dynamical state of the laser. In addition, we will study how the mirror reflectivity influences the excitable state.

The rate-equation model and underlying compact models will be explained first. The model will then be used to present the evolution of gain and absorber carrier densities and photon concentrations in phase portraits. Subsequently, we will show how for four different combinations of gain current and reverse-bias voltage the laser shows qualitatively very different dynamical states. Furthermore, we will use the model to determine for which mirror reflectivity values, the excitable control parameter space is largest.

## 2. Laser structure

The modelled laser structure is a two-section laser comprising a saturable absorber (SA), semiconductor optical amplifier (SOA), and passive elements formed by mirrors $R_1$ and $R_2$ as depicted in Figure 1 (bottom). Such a modular approach is compatible with the commercially available multi-project wafer (MPW) InP platform discussed in [24]. The optical cavity is formed by combining various building blocks such as a broadband multimode interference reflector [25] (MIR) and a distributed Bragg reflector [26] (DBR). As indicated in Figure 1, an external perturbation in the form of an optical trigger pulse can be injected into the cavity from the MIR side of the laser, while the photon number in the cavity can be related to the optical output on the DBR side. The active regions of both SOA and SA consist of four multi-quantum wells (MQW) to provide optical absorption and gain by applying either a reverse-bias voltage or forward current, respectively. The different building blocks are connected using ridge waveguides with the cross-section shown in Figure 1 (top) with a total length indicated by the passive element in Figure 1 (bottom). Electrical contacts are deposited on the top and bottom of the layer stack for electrical probing.

The spectral absorption [27] and gain parametrisation [28], [29] of the SA and SOA building blocks were studied before, and experimental data are used in this work to obtain functional relations for the rate-equation model.

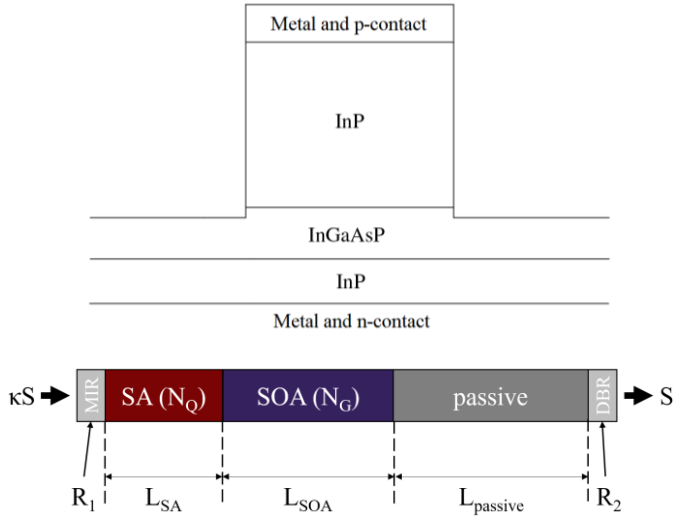

*Figure 1: Top: Ridge waveguide cross-section in the generic platform [24]. Bottom: Two-section laser comprising two mirrors, a saturable absorber (SA), semiconductor optical amplifier (SOA), passive elements, and in- and output.*

## 3. Method

*Spiking Laser Model*

The model under investigation is a lumped-cavity description [22], [30] comprising three coupled differential equations to describe the gain and absorber carrier densities $N_G$ and $N_Q$ and the total photon number $S$ in the cavity. The mathematical descriptions of the gain and absorber region are based on parametrisations of the gain, carrier lifetime and other parameters of the InP platform using compact models.

$$\frac{dS}{dt} = v_g g_{net} S + v_g \alpha_{sat} S - \frac{S}{\tau_{ph}} + S_{sp} + \kappa s(t) \quad (1)$$

$$\frac{dN_Q}{dt} = -v_g \alpha_{sat} \frac{S}{V_{SA}} - \frac{N_Q}{\tau_Q} \quad (2)$$

$$\frac{dN_G}{dt} = \frac{I}{eV_{SOA}} - v_g g_{net} \frac{S}{V_{SOA}} - \frac{N_G}{\tau_G} \quad (3)$$

For a listing of the parameter names and values, we refer the reader to Table 1. Equation (1) describes the rate at which photons are generated. The terms in the right-hand side correspond, in the order of appearance, to the SOA gain $g_{net}$ (stimulated emission), the absorption in the SA $\alpha_{sat}$, the cavity loss expressed as the photon lifetime $\tau_{ph}$, the spontaneous emission $S_{sp}$ in the cavity and the injection of external photons in the cavity through $\kappa s(t)$. Equation (2) describes the absorber carrier density rate of change, in which the first term gives the photon-induced change in the absorber volume $V_{SA}$ and the last gives the carrier-density loss expressed in the absorber carrier lifetime $\tau_Q$. Equation (3) models the gain carrier density, where the first term accounts for injected carriers through current $I$ into gain volume $V_{SOA}$. The second term describes reduction of the carrier density due to optical gain, and the last term models the gain carrier recombination rate of the gain through lifetime $\tau_G$.

In (1), the net gain $g_{net}$ is based on a logarithmic expression in the gain carrier density $N_G$ [31] with respect to the gain transparency carrier density $N_{G0}$, and takes into account the material gain and passive loss per unit length.

$$g_{net} = \Gamma \left[ a_{ng} N_{G0} \log \frac{N_G}{N_{G0}} \right] - \left[ c_1 \left( \frac{N_G}{N_{G0}} \right)^2 + c_2 \frac{N_G}{N_{G0}} + c_3 \right] \quad (4)$$

Here, $\Gamma$ is the optical confinement factor, and $a_{ng}$ the gain cross-section for stimulated emission, both of which contribute to the material gain. The passive loss was previously measured on the generic platform for current densities ranging between 1.0 to 10.0 kA/cm² and modelled using a quadratic function with fitting parameters $c_i$ ($i$=1,2,3) of the carrier density ratio $N_G/N_{G0}$ [32].

The absorption includes a saturation effect as shown in Equation (5), where $\alpha_{sat}$ is proportional to the differential absorption $a_{nq}$ and the absorber carrier density $N_Q$ with respect to the absorber transparency carrier density $N_{Q0}$. The saturation of the absorption is accounted for by the saturation photon number $S_{sat}$, which corresponds to a saturation energy of 1 pJ, assumed to be valid for quantum well-based absorbers [33].

$$\alpha_{sat} = \frac{\Gamma \left[ a_{nq} (N_Q - N_{Q0}) \right]}{1 + \frac{S}{S_{sat}}} \quad (5)$$

The photon lifetime $\tau_{ph}$ is mainly determined by the mirror losses and is defined by [34]:

$$\tau_{ph} = \frac{1}{v_g \alpha_{mirror}} \quad (6)$$

where mirror losses $\alpha_{mirror}$ are influenced by the structural parameters of the laser, namely the mirror reflectivities $R_1$ and $R_2$ and the total cavity length [35]:

$$\alpha_{mirror} = \frac{\ln \left( \frac{1}{R_1 R_2} \right)}{L_{SA} + L_{SOA} + L_{pass}} \quad (7)$$

The spontaneous emission rate $S_{sp}$ depends on the gain carrier density $N_G$, as well as the bimolecular recombination rate $B$, and the spontaneous emission factor $\beta$, which relates the number of spontaneously generated photons coupled into the lasing mode [4]:

$$S_{sp} = V_{gain} \beta B N_G^2 \quad (8)$$

The photon number $S$ in the cavity translates to the optical power via [4]:

$$P_{\text{opt}} = \frac{S h c_0 \eta_c}{\tau_{\text{ph}} \lambda} \quad (9)$$

in which $hc_0/\lambda$ is the single-photon energy, and $\eta_c$ an outcoupling efficiency which depends on the mirror reflectivity. Vice versa, an estimation of the number of photons in an external pulse injected into the cavity can be calculated from Equation (9):

$$S_{\text{inj}} = \frac{P_{\text{opt}} \tau_{\text{ph}} \lambda}{h c_0 \eta_c} \quad (10)$$

*Compact models*

In the rate-equation formula that describes the absorber (Equations (2)), the voltage dependency is modelled using a voltage-dependent absorber carrier lifetime $\tau_Q$ and saturation factor $\alpha_{\text{sat}}$ via a voltage dependent absorber transparency carrier density $N_{Q0}$. The model that estimates the absorber carrier lifetime $\tau_Q$ is based on previously measured carrier sweep-out times in an electro-absorption modulator (EAM) on the same technology platform [36]. The carrier lifetime follows an exponential function depending on the applied reverse-bias voltage:

$$\tau_Q = a \cdot e^{b \cdot V_{\text{rb}}} \quad (11)$$

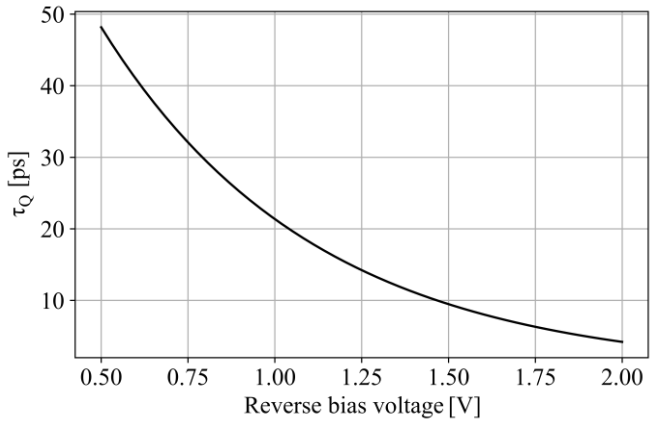

*Figure 2: Extracted absorber carrier lifetime as a function of the applied reverse-bias voltage for a = 1.086·10⁻¹⁰ s, and b = -1.626 V⁻¹.*

Since the absorber transparency carrier density is voltage dependent [37], data obtained previously from small-signal absorption measurements $Q_0$ for different reverse-bias voltages of a saturable absorber are used [33] to create a compact model for the transparency-versus-voltage relationship. Analytically, the small-signal absorption can be expressed as [38]:

$$Q_0 = \Gamma a_{\text{nq}} \left( N_{Q0} - \frac{I_A \tau_Q}{eV} \right) \quad (12)$$

where $I_A$ is an applied direct injection current. Thus, any variation in $N_{Q0}$ changes the small-signal absorption. Without additional current injection, i.e., $I_A = 0$ mA, the small-signal absorption is reduced to (see also [33], [39]):

$$Q_0 = \Gamma a_{\text{nq}} N_{Q0} \quad (13)$$

Measurements of the optical transmission through a 90 μm saturable absorber under different reverse biases were used to determine the relation between $Q_0$ and $V_{\text{rb}}$. Equation (13) was then used to match $N_{Q0}$ to the different reverse biases, assuming the absorber transparency carrier density at $V_{\text{rb}}=0V$ and $\lambda=1560$ nm is $0.05 \cdot 10^{24}$ m⁻³. This value was obtained before in pulse-transmission measurements and matching carrier density simulations of a saturable absorber on a similar InP-based technology platform [33]. In Equation (13), $a_{\text{nq}}$ and $N_{Q0}$ are matched to the small-signal measurements, which yields the linear relationship between absorber transparency carrier density and reverse-bias voltage at 1550 nm depicted in Figure 3.

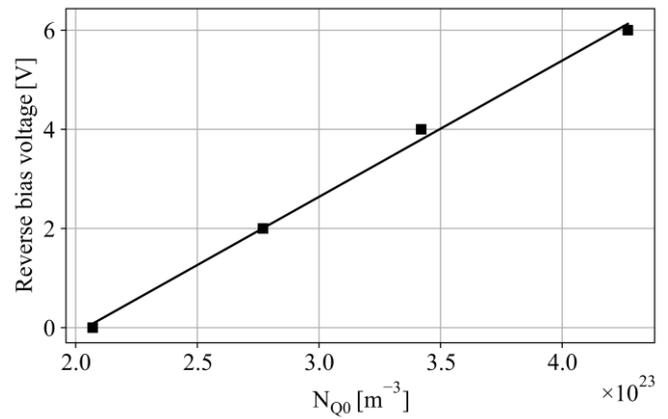

*Figure 3: Reverse-bias voltage $V_{rb}$ as a function of transparency carrier concentration $N_{Q0}$ at 1550 nm. Black squares: estimated values of $N_{Q0}$ at 1550 nm. Solid line: linear fit.*

*Solving the rate-equation model numerically*

The rate-equation model Eq.(1)-(3), logarithmic and linear gain and absorber models Eq.(4)-(5), and underlying compact models Eq.(6)-(11) were solved numerically using the Adams–Bashforth integration method implemented in Python to find solutions for $S$, $N_Q$ and $N_G$. Unless stated otherwise, for all simulations the values mentioned in Table 1 are used.

*Table 1: Simulation parameters. Var. indicates the actual parameter value is based on other control or design parameters.*

| Photon lifetime | $\tau_{\text{ph}}$ | 31.3 | ps |
|---|---|---|---|
| Gain carrier lifetime [28] | $\tau_G$ | 300 | ps |
| Absorber carrier lifetime | $\tau_Q$ | var. | ps |
| Group velocity | $v_g$ | $8.2 \cdot 10^7$ | m·s⁻¹ |
| Spontaneous emission rate | $S_{\text{sp}}$ | var. | s⁻¹ |
| Gain cross-section | $a_{\text{ng}}$ | $1.75 \cdot 10^{-19}$ | m² |
| Differential absorption | $a_{\text{nq}}$ | $4.00 \cdot 10^{-19}$ | m² |
| Gain transparency carrier density | $N_{G0}$ | $5.00 \cdot 10^{23}$ | m⁻³ |

| Absorber transparency carrier density [33] | $N_{Q0}$ | var. | $m^{-3}$ |
|---|---|---|---|
| Confinement factor | $\Gamma$ | 0.053 | [-] |
| Gain medium volume | $V_{SOA}$ | $1 \cdot 10^{-16}$ | $m^3$ |
| Absorber medium volume | $V_{SA}$ | $1 \cdot 10^{-17}$ | $m^3$ |
| Bias current | $I$ | var. | A |
| Optical injection rate | $\kappa_S(t)$ | var. | $s^{-1}$ |
| Coupling efficiency | $\eta_c$ | 0.6 | [-] |
| Lasing wavelength | $\lambda$ | 1550 | nm |
| Absorber reverse-bias voltage | $V_{rb}$ | var. | V |
| Spontaneous emission factor | $\beta$ | $1 \cdot 10^{-4}$ | [-] |
| Bimolecular recombination rate | $B$ | $1 \cdot 10^{-16}$ | $m^3 s^{-1}$ |
| DBR mirror reflectivity | $R_1$ | 0.866 | [-] |
| MIR mirror reflectivity | $R_2$ | 0.632 | [-] |
| Absorber section length | $L_{SA}$ | 50 | μm |
| Gain section length | $L_{SOA}$ | 500 | μm |
| Passive section length | $L_{pass}$ | 1000 | μm |

## 4. Results

Since the model at hand defines a three-dimensional system of equations, the numerical solutions can be visualized as a trajectory in phase space. Depending on the initial conditions at t=0, laser parameters and external optical perturbation condition, the solution can evolve, but need not, to a stable limit cycle, representing a steady state of the system. In 3D, there is also a possibility that no limit cycle exists, i.e., the cycle period is infinite in time. A chaotic attractor is an example of the latter.

First, a laser with the parameters given in Table 1, a gain current of 40.5 mA and reverse-bias voltage of 0.6 V is simulated for a time period from 0 to 3 ns. An optical pulse with a Gaussian profile, time duration of 50 ps, and peak power of 5 mW is injected at t=1 ns. Figure 4 shows the evolution (solid black line) as a trajectory in 3D phase space. The three light-grey lines are the projections onto the $(N_G, S)$, $(N_Q, S)$, and $(N_G, N_Q)$ 2D phase planes. The black dot and square denote the initial condition (simulation starting value), and simulation end point, respectively. The moment at which the optical pulse is injected is highlighted with a black triangle. From the simulation starting point, the solution slowly drifts towards a higher value for $S$ and $N_Q$, and a lower value in $N_G$. A large excursion of $S$ follows, before the solution moves towards the steady state with a very small value for $S$. When the external optical pulse is injected, a rapid increase in $S$ and decrease in $N_G$ follows. The result is a second but slightly larger excursion in phase space. The difference between the two excursions originates from the different triggering mechanisms. The first excursion is the result of the initial conditions, whereas the second excursions is the result of a larger number of photons injected into the simulated cavity.

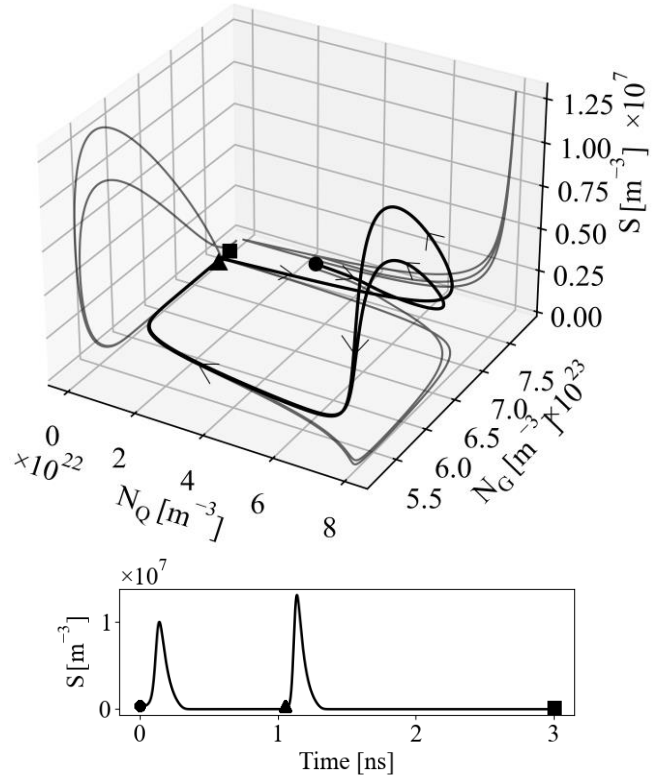

*Figure 4: Numerical solutions of an excitable laser under optical injection in three dimensional phase space (top, solid black line). The grey lines are projections of the 3D solution onto 2D phase planes. ● simulation start point, ■ simulation end point, ▲ pulse injection. Bottom: corresponding time trace.*

### Tuning laser control parameters

In the previous examples, the laser was biased in a specific operation point to show its dynamics and excitable behaviour. However, depending on the combination of different gain currents, reverse biasing voltages and optical pulse injection, the laser can show different operation modes. Figure 5-8(a)-(d) show four different operation regimes and corresponding two dimensional phase space plots when laser control parameters (gain current and reverse-bias voltage) are changed.

First, in Figure 5, the laser is operated at a gain current of 87.1 mA and reverse-bias voltage of 1.800 V and no optical pulse is injected. From the simulated time trace (Figure 5(d)) clearly the laser shows a sustained pulsed output with a fixed repetition rate. In $(S, N_G)$ and $(N_G, N_Q)$ phase space, this is visible as an asymptotic stable limit cycle: for every generated pulse, the same closed loop is followed. From a dynamical perspective, this behaviour represents a periodically spiking neuron or neuron in burst mode [6].

Close to the self-pulsating operation point exists a dynamically very different state. When the reverse-bias voltage is changed by only 3 mV while the gain current remains the same, the laser suddenly loses the capability to sustain a pulsed output, instead it shows damped oscillations before reaching a stable optical cw output (Figure 6(d)). These relaxation oscillations are fast, since their frequency is determined by the fast photon lifetime and the resonant energy exchange between the optical field and population inversion [18], [22]. In phase space, the limit cycle has disappeared and is replaced by a spiral into a stable fixed point (Figure 7(a)-(c)).

Next, the gain current is lowered to 55.0 mA, the reverse-bias voltage is changed to 1.000 V, and an optical trigger pulse is injected at t=1.1 ns. From Figure 7(d) it is clear that the trigger caused the system to start lasing in cw mode. At the moment the pulse is injected, the output is raised to a steady state. Apparently, the system is bistable, i.e., on or off. Qualitatively, the dynamical behaviour is similar to that presented in Figure 6, albeit with a different relaxation oscillation frequency, consistent with the smaller photon number.

Lastly, the gain is kept at 55.0 mA, but the reverse-bias voltage is set at 1.500 V. Figure 8(d) shows the laser is now excitable, since the optical pulse now triggers the laser to generate a single pulse. The observed dynamics of this excitable laser neuron is analogous to excitability observed in biological neurons [6], [40].

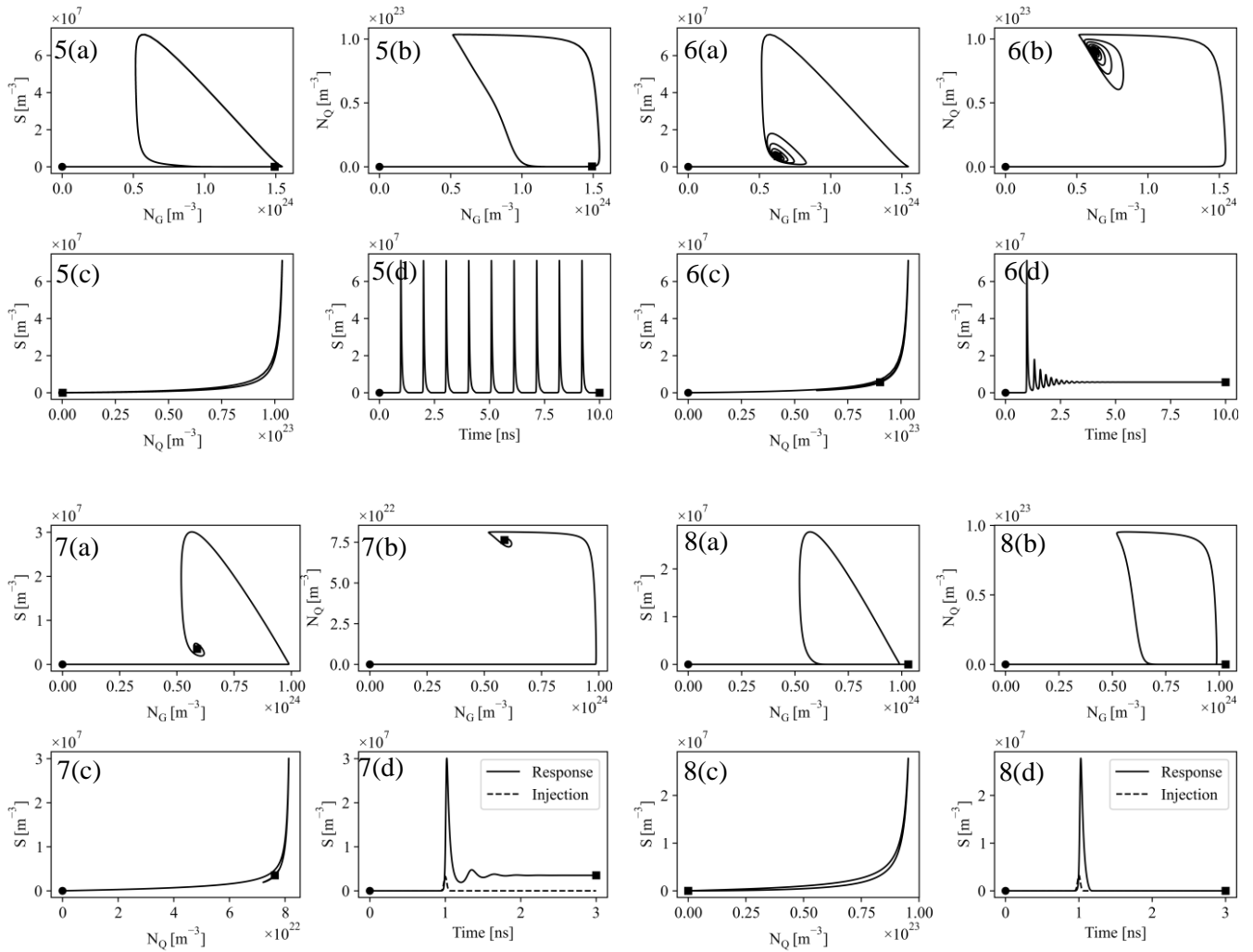

*Figure 5-8: Transients in four operation regimes of the simulated laser. Fig. 5: from off to self-pulsating, Figs. 6 and 7: from off to cw, Fig. 7: trigger-induced switch on, Fig. 8: excitability. In Fig.5 and 6, the gain current is fixed at 87.1 mA, while in Fig. 5, the reverse-bias voltage is 1.800 V, and in Fig. 6 1.797 V. In Fig. 7-8, the current is lowered to 55.0 mA, while the voltage is 1.000 V and 1.500 V, respectively. In these cases, an optical pulse with width of 50 ps, and a peak power of 5 mW is injected at t=1.1 ns.*

*Tuning laser design parameters*

The four cases described previously validate that our model is capable of generating different laser operation states. Another key aspect to investigate is the influence of laser design parameters on the dynamical states of the laser neuron. One design parameter of interest is the reflectivity of one of the cavity mirrors, which can be chosen when designing a two-section laser for tape-out. The reflectivity of the DBR mirror in Figure 1 can be altered by changing its length [26]. The reflectivity of the MIR mirror is fixed by design. In order to investigate the effect of different reflectivities on the laser operation states, we first use the model to calculate trajectories at a fixed gain current and four different reverse-bias voltages. In Figure 9(a), and (c), the simulation results of two different laser designs are depicted in 3D phase space. The trajectory starting and end points are indicated by the black circle and square, respectively. The corresponding time traces are depicted in Figure 9(b) and (d). In both simulations, the gain current was set to 50 mA, while the reverse-bias voltage was swept from 0 to 3 V in increments of 1 V. An external optical trigger pulse with a pulse width of 50 ps and peak power of 5 mW is applied at t=1.0 ns, indicated by the dash-dotted lines in Figure 10(b), and (d).

In the first case, Figure 9(a)-(b), the structure of the laser is the same as in the previous simulations, which means that the DBR-mirror reflectivity is set to be 0.866. At a reverse-bias voltage of 3 and 2 V (black lines), the laser response is below the excitable threshold. In Figure 9(a), this is clear from the small loops and the low value for $S$. At these reverse-bias voltages, the absorption is high. When the reverse-bias voltage is lowered to 1 and 0 V, absorption decreases and large excursions in phase space and pulses in the time traces due to the injected pulse are observed, indicating the laser is excited. This is indicated by the red lines in Figure 9(a), and (b). In Figure 9(c), and (d), the reflectivity of the DBR mirror was lowered to 0.400. Effectively, this changes the loss in the cavity, which influences the photon lifetime, as indicated by Equation (6) and (7). Consequently, the dynamical states change for the same applied reverse-bias voltage as in the previous case. By inspecting Figure 9(c)-(d), it is observed that at a reverse-bias voltage of 3 V, the laser trajectory shows the same small excursions (black line). However, at a reverse-bias voltage of 2 and 1 V a large excursion is observed (red). Thus, the laser is excitable at these voltages. Moreover, for the case where the reverse-bias voltage is 0 V (blue), the laser does not follow a closed trajectory, but moves to a stable fixed point, resulting in the cw operation mode. In this case, the cavity losses are lower due to a higher value for the reflectivity, therefore the laser operates in cw. In the previous example, the laser was excitable for these parameter settings.

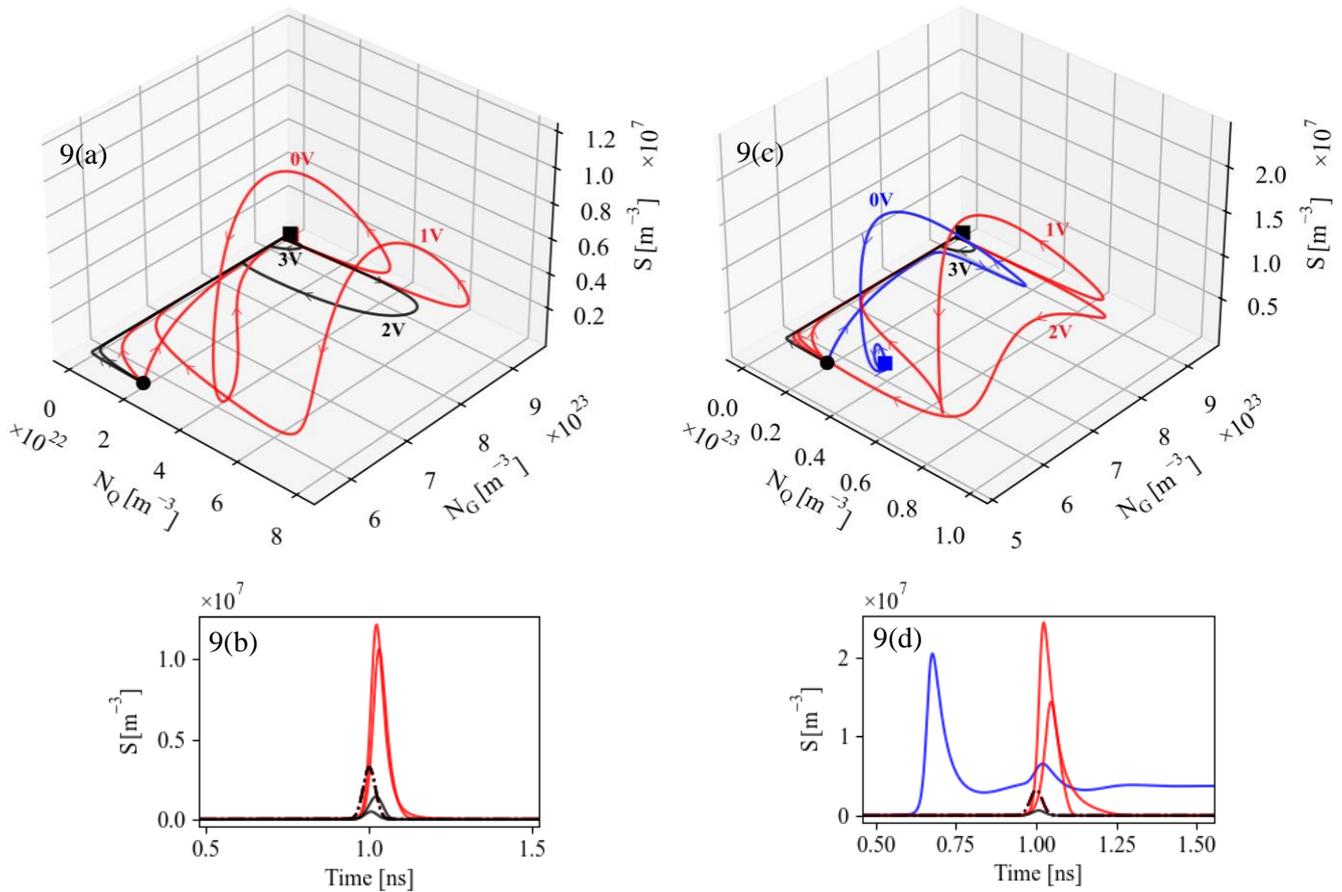

*Figure 9: Simulation results of two different laser designs. Top figures (a), (c) show the trajectories in 3D phase space for reverse bias voltages of 0, 1, 3, and 4V. Bottom figures (b), and (d): simulated time traces showing the injected pulse at t=1.0 ns (dash-dotted line) and laser response (solid line) Left case: mirror reflectivity 0.866, right case: mirror reflectivity 0.400.*

The two cases discussed in Figure 9 indicate qualitatively that the laser control and design parameters influence the dynamical state the laser operates in. Depending on the mirror reflectivity, the laser may or may not be excitable under optical-pulse injection. Next, we perform a control and parameter sweep in order to get qualitative results on the excitable regime. More precisely, we investigate the dependency of the gain current, reverse-bias voltage and the mirror reflectivity on the excitable operation regime.

First, In Figure 10(a)-(d), the simulation results of four different laser designs with DBR reflectivity values of 0.3, 0.5, 0.7, and 0.9 are shown, which could be achieved by changing the DBR length to 0.15, 0.20, 0.30 and 0.35 mm [26]. In these simulations, the gain current was swept from 30 to 90 mA, and for every value for the gain current, the reverse-bias voltage was swept from -1 to 4 V. An external optical trigger pulse with a pulse width of 50 ps and peak power of 5 mW is applied at t=4.0 ns. For every simulation point, the average optical output power is calculated over a simulation window of 25 ns. Based on the calculated optical output, the operation mode of the laser was determined. The parameter space for excitable operation is depicted in Figure 10(a)-(d).

By comparing the four individual maps, it is clear when the laser is biased at a fixed gain current, the largest excitability parameter space is obtained if the mirror reflectivity is low. Gradually increasing the reflectivity from R=0.3 to R=0.9 (Figure 10(a) to (d)) results in a reduction of the parameter space for excitable operation. For example, for a reflectivity of R=0.3 excitable operation at a gain current of 50.0 mA was observed between a reverse-bias voltage of -0.175 and 1.17 V ($\Delta_{Vrb}$=1.34 V), whereas changing the mirror reflectivity to R=0.9 results in excitable operation between 1.00 and 1.83 V ($\Delta_{Vrb}$=0.830 V). This is highlighted with the red vertical line in Figure 10(a) and (d). Similar results are observed when considering a the gain current. At a fixed reverse-bias voltage of 1.00 V, excitability is observed between 48.0 and 73.6 mA ($\Delta_I$=25.6 mA) for the case where R=0.3, whereas increasing the reflectivity to R=0.9 results in excitable operation between 38.0 and 53.3 mA ($\Delta_I$=15.3 mA), as indicated by the horizontal red line. Decreasing the reverse-bias voltage to 0.140 V reduces the excitable operation space to $\Delta_I$=13.7 mA and $\Delta_I$=4.02 for R=0.3 and R=0.9, respectively.

To further quantify this observation, two simulations at reverse-bias voltages of 0.140 V and 1.00 V for mirror reflectivities swept from near 0 to 1 and gain currents between 20.0 and 90.0 mA were performed. In Figure 10(e) and (f), the parameter space for excitable operation for different values of the reflectivity and gain currents are depicted. By comparing the area of excitable operation in these two maps, it is clear that the largest parameter space for excitable operation exists for relatively low reflectivity values (i.e. R < 0.5). At higher reflectivity values, the excitability parameter space decreases, which is consistent with the observation in Figure 10 (a)-(d). Also, when the reflectivity approaches zero, excitability vanishes, due to the cavity losses. In addition to the area of excitability, the average optical output power of the excited pulse is mapped onto a colour scale. In both cases, it is clear that at the left and right boundary of excitability, the optical output power is lowest and highest, respectively. Figure 10(e) and (f) are especially of interest when the total laser energy consumption is considered. By operating the laser at the lowest possible gain current, while ensuring a relatively large window of excitable operation is achieved, static power consumption is minimal.

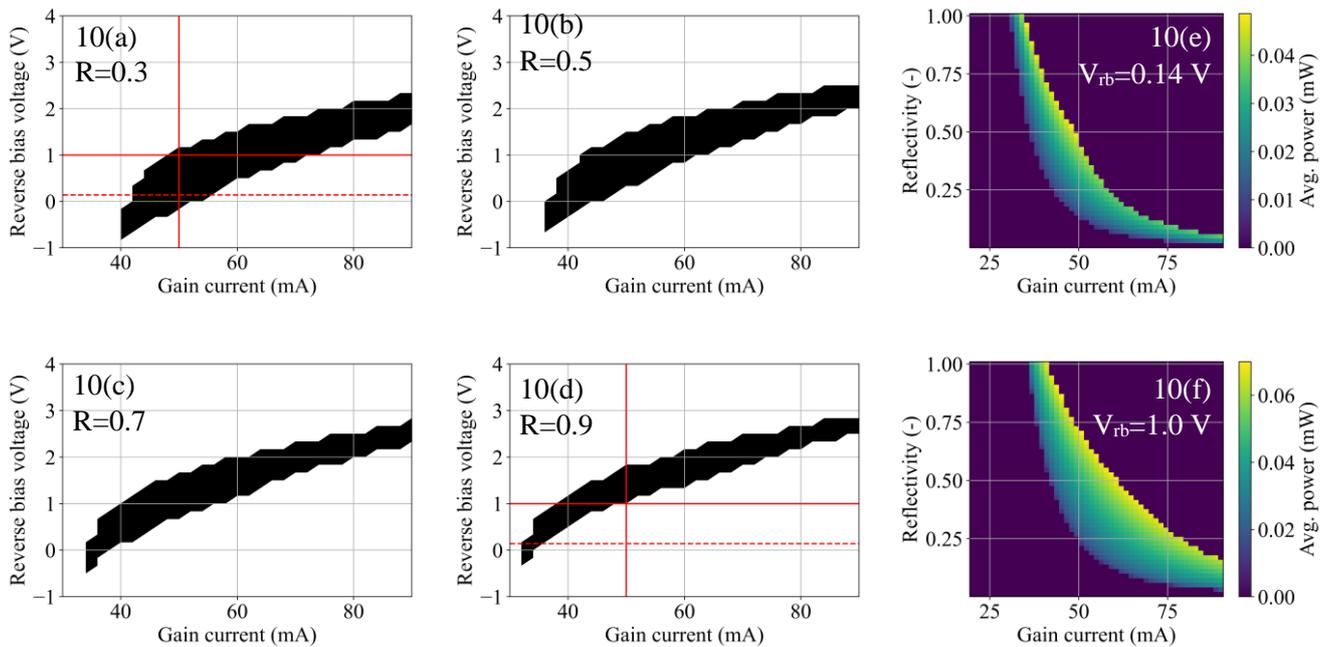

*Figure 10 (a)-(d): Simulation results of four different laser designs. The reflectivity of the DBR mirror is swept from 0.3 (a) to 0.9 (d). Reflectivity of the MIR is fixed at 0.6. Other simulation parameters remain as described in Table 1. Figure 10 (e)-(f): area of excitable operation for $V_{rb}$=0.14V (e) and 1.0V (f). The colour bars indicate the total optical output power of the excited pulse.*

## 5. Conclusion

In this work, we used a rate-equation model and compact models based on the generic InP technology platform to numerically calculate the evolution of gain and absorber carrier densities and photon concentration. This allowed us to study the dynamics of an integrated two-section laser under optical pulse injection. Using these models, the operation mode and evolution of the dynamics for two control parameters, i.e. the gain current and reverse-bias voltage, can be determined.

From the phase-portrait analysis, a classification of the laser operation modes is presented. First, at a gain injection current and reverse-bias voltage of 87.1 mA and 1.8 V, respectively, the model shows a self-spiking mode where the solution follows a limit cycle in $(S, N_G, N_Q)$ phase space with sustained oscillations in the calculated time-trace. The limit cycle is replaced by a stable spiral when the reverse-bias voltage is changed by only 3 mV, indicating the laser pulses are replaced by fast relaxation oscillations before cw mode is reached. When the gain injection current is lowered to 55.0 mA and an optical trigger pulse is injected, the on-set mode and excitable mode are predicted.

Besides the control-parameter dependent dynamical state, we further demonstrated that depending on the value for the cavity-mirror reflectivity, the laser shows different trajectories in three dimensional phase space. The main reason for this are the mirror-reflectivity dependent cavity losses and photon lifetime, indicating that, in order to operate the laser as an excitable integrated laser neuron, the value for the reflectivity should be carefully chosen. We presented simulation results where the mirror reflectivity was swept from near 0 to 1 for two different reverse-bias voltages of 0.140 V and 1.00 V. In addition, the optical output of the excited pulse was recorded. The results demonstrate that the largest parameter space for excitable operation is obtained for relatively low reflectivity values. Moreover, output-pulse energies are lower at smaller gain currents. This observation is of great importance when considering low energy consumption, i.e. a low gain current operation.

## Data availability statement

All data that support the findings of this study are included within the article (and any supplementary files). All data are available in the manuscript or the supplementary materials.

## Acknowledgements

The authors would like to acknowledge Dr. Erwin Bente for providing the fit model for the passive losses and data on the transparency carrier density, Dr. Martijn Heck for providing data for compact models, and Dr. Robert Otupiri for the discussions on the phase space results.

## Author Contributions

Lukas Puts: performed all simulations, revised model, implemented model in Python, interpretated results and drafted manuscript, prof.dr. Daan Lenstra: supported in phase-space analysis, revised model, interpretated results and revised manuscript., prof.dr. Kevin Williams: revised manuscript, and dr. Weiming Yao: developed model, implemented model in Python, supported interpretation and analysis of results, revised manuscript.

## Conflict of Interest

The authors declare no conflict of interest.

## ORCID iDs

Lukas Puts https://orcid.org/0000-0001-5416-3006
Daan Lenstra https://orcid.org/0000-0002-4000-8897
Kevin Williams https://orcid.org/0000-0001-9698-9260
Weiming Yao https://orcid.org/0000-0002-4558-317X